# *In situ* TEM observation of oxygen evolution reaction in GDC-Pt half cell at elevated temperatures

Amir Hossein Tavabi[1], Takayoshi Tanji[2]

[1]Ernst Ruska-Centre for Microscopy and Spectroscopy with Electrons and Peter Gruenberg Institute, Forschungzentrum Juelich, Juelich, Germany

[2]Ecotopia Science Institute, Nagoya University, Furo-cho, Chikusa, Nagoya 464-8603, Japan

*Abstract*

*Gadolinium-doped ceria was studied in a half SOFCs at elevated temperatures. In situ electron holography could reveal ionic activity in a single GDC-Pt interface at the elevated temperatures that result in further oxidation of the GDC. In situ EELS measurements confirmed holography results and showed the existence of the gadolinium in a new oxidation state. Also, it was found that the doped cations are the active ions in the oxygen evolution reaction from the solid oxide electrolyte.*

## 1. Introduction

Gadolinium-doped ceria (GDC) is an important electrolyte material for solid oxide fuel cells (SOFCs) that shows both ionic and electronic conductivity [1]. In this oxide, gadolinium cations ($Gd^{3+}$) will substitute in the ceria fluorite structure with a lower oxidation state comparing with cerium cations ($Ce^{4+}$), and hence introduce oxygen vacancies into the cubic structure and improve the oxygen ions conductivity as the electrolyte material of the SOFCs [2].

Some studies of in situ transmission electron microscopy (TEM) on the electronic and crystal structures of ceria and ceria-zirconia solid solutions have shown that $Ce^{4+}$ can be easily reduced to $Ce^{3+}$ in a reducing environment [3,4]. This phenomenon, which was also referred to as the oxygen release/absorption effect of ceria [5], has an important influence on the catalytic activity of several metal catalysts supported on ceria [6,7] or ceria-based solid solutions [8-10]. There are a few reports on the electronic structure of GDC at room temperature [11-13]. It has been believed that in an SOFC with GDC as the electrolyte, the cerium cations, which may undergo reduction on the anode side, are mainly responsible for the ionic conductivity, while the doped cations simply serve to facilitate the ionic conduction through the oxygen vacancies [14,15]. However, to the best of the authors' knowledge, there has been no in situ analytical electron microscopy study of GDC used as an SOFC electrolyte material.

Among various analytical microscopy techniques off axis electron holography offers a superior precision to detect fine changes or evolutions in the specimen [16]. This technique is able to resolve the electric potential distribution in the object that can be interoperated as the ionic activity in in situ investigations [17, 18].

It the present work in situ electron holography and in situ electron energy loss spectroscopy (EELS) techniques were applied to the study of GDC as the electrolyte material in a half solid oxide cell.

2. *Materials and methods*

A half-cell of an SOFC was fabricated on a Si single crystal substrate by pulsed laser deposition (PLD). A thin Pt layer of 70nm was deposited on the substrate as the half cell electrode at room temperature and in a high vacuum chamber ($1\times10^{-6}$ Pa). The GDC electrolyte layer with 20mol% gadolinia was deposited on the electrode at 400°C and in $4\times10^{-2}$ Pa oxygen atmosphere. An X ray diffraction pattern of the deposited electrolyte layer showed the typical structure of the GDC fluorite structure.

Cross-sectional specimens for in situ TEM observations were prepared by focused ion beam (FIB) micro-sampling technique using Hitachi FB2100. The half-cell structure was characterized by energy dispersive x-ray spectroscopy (EDS) technique using a JEOL ARM200F TEM. Figure 1 shows the line profile of the structure components along the perpendicular direction to the electrode-electrolyte interface.

In situ off axis electron holography observation were performed by a cold field emission TEM, Hitachi HF-2000, operating at 200kV. This TEM has been equipped with an electron biprism close to the image plane of the objective lens to make holograms. In off axis electron holography a so called object electron wave that has been transmitted through the observed specimen includes the information of the specimen. Such information in the objective wave can be recorded by interfering with a reference wave which has been passed just through vacuum to make a hologram. Subsequently, this hologram must be reconstructed to acquire the phase and amplitude images. In the specific condition the phase of the electron wave which is shown as the phase image can be directly related to the inner electrostatic potential of the observed object. Therefore it is possible to monitor and understand the change in the specimen during the in situ observations.

In situ EELS measurements were conducted by the high voltage electron microscope JEM 1000K RS equipped with a Gatan GIF Quantum. The microscope was operated at 800kV in the present experiment. An aperture whose diameter corresponded to 74nm in the objective plane was used in GIF in order to limit the area, where the information was obtained from, to the desirable areas.

*3. Results and discussion*

Figure 2a shows the half-cell structure, which includes the Si substrate, the Pt electrode, the GDC electrolyte, and a W layer deposited by FIB to protect the GDC layer from ion beam damages during the specimen preparation.

Energy loss spectrometry measurements were conducted at room temperature and 400°C on the Pt/GDC interface, which had catalytic properties, and on the GDC/W interface, which had no catalytic properties. The exact areas of the interfaces from where the spectra were acquired in the both environments are shown Figs. 2b and 2c.

The cerium and gadolinium energy loss spectra obtained from the GDC/W interface at room temperature and 400°C are shown in Figs. 3a and 3b. No considerable change in the spectra was observed during the heating, implying that no reaction occurred at this interface. In contrast, the spectra obtained from the Pt/GDC interface at 400°C showed an abnormal change. Interestingly, the electronic state of ceria in Pt/GDC interface showed a steady state even at elevated temperatures (Fig. 4a). However, there was a significant difference between the gadolinium energy loss spectra at room temperature and 400°C (Fig. 4b), where chemical shifts of 3 eV could be found at the M5 and M4 edges towards the higher-energy side, and this was accompanied by a considerable increase in the relative intensity of the M5 white line. This change suggests the existence of the gadolinium cations in a new oxidation state ($Gd^{4+}$) in the stable ceria matrix.

It is well known, that the reduction of $Ce^{4+}$ to $Ce^{3+}$ would bring about a clear reversal in the relative intensities of the M5 and M4 lines, as well as a change in the shoulder of the M5 andM4 white lines and the concomitant chemical shifts to the lower-energy side of the spectrum [3,8,10]. Therefore, the observed chemical shifts are confirmed to be due to the change in the gadolinium valence state and not the cerium oxidation state.

These results shows that during the heating process, while the oxygen ions were released from the GDC solid oxide through the Pt interface, as per the reaction

$$2O^{2-} \rightarrow 4e^{-} + O_2,$$

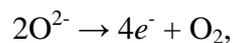

both the charge compensation in the electronic-ionic conductor (GDC) and the charge transfer to the grounded Pt electrode were occurred.

The half-cell was also studied by in situ off axis electron holography. Holograms were recorded at different temperatures and were reconstructed to acquire the phase images. The phase profiles were extracted from the phase images along to the longer side of the white rectangle in the Fig. 2a, averaged along the shorter side. Figure 5 shows the phase shift profiles at room temperature, 200°C, and 300°C. The phase sign in these profiles was selected as negative, that is, it corresponds to the electrostatic potential for the electrons. Therefore, the vertical axes in these profiles represent the inner potential distribution in the half cell at every temperature. However, since the absolute phase value of the Pt electrode is higher than the GDC electrolyte, the scale of the vertical axis (radian) was set so that the electrode-electrolyte interface to be zero in order to simplify the comprehension.

As shown in Figs. 5b and 5c, by increasing the temperature of the cell no considerable change can be distinguished at the W/GDC interface that means there was not any ionic activity in this interface. However, an obvious negative shifts can be seen in the phase values of the electrolyte close to the Pt interface, in words, the difference of the phase amounts between the Pt electrode and the GDC electrolyte was reduced as the temperature increased. This phenomenon shows that the inner potential of the GDC was made deeper during the heating process, that is the oxide was positively charged and the material was gone under further oxidation.

These results are in complete agreement with in situ energy loss spectroscopy measurements and confirm that the ceria was not reduced at the elevated temperatures and the gadolinium cations were more oxidized during the oxygen ion transferring at the Pt interface.

## 5. Conclusion

GDC was studied as the electrolyte of a half cell by two in situ analytical TEM approaches at the elevated temperatures. The results of both techniques were consistent and denoted that the GDC experiences a higher state of oxidation at higher temperatures.

Also, the results suggested that doped cations play an important role in ionic conduction and the mechanism underlying of the redox reactions in the SOFCs with a mixed ionic-electronic conductor electrolytes such as GDC. That is, the gadolinium cations are responsible cations for the reactions or at least will trigger the reactions.

This phenomenon is the reverse of that was observed in an electronically insulator ionic conductor such as yttria-stabilized zirconia (YSZ) in which the oxide electrolyte goes under a reduction during the heating process and the oxidation state of the zirconium cations, as the host cations, is reduced to a lower oxidation state [17,18].

However, in situ off axis electron holography showed a higher sensitivity compare the other analytical technique and could reveal the reaction at the relatively lower temperatures.

*References*

[1]  Steele, B. C. H.; Solid State Ionics (129) 2000 95-110.

[2]  Mogensen, M.; Sammes, N.M.; Tompsett, G.A. Solid State Ionics (129) 2000 63-94.

[3]  Crozier, P.A.; Wang, R.; Sharma, R. Ultramicroscopy (108) 2008 (1432-1440.

[4]  Gao, P.; Wang, Z.; Fu, W.; Liao, Z.; Liu, K.; Wang, W.; Bai, X.; Wang, E. Micron 41 (2010) 301-305.

[5]  Arai, S.; Muto, S.; Sasaki, T.; Ukyo, Y.; Kuroda, K.; Saka, H. Electrochem. Solid-State Lett. 9 (2006) E1-E3.

[6]  Wang, R.; Crozier, P.A.; Sharma, R. J. Phys. Chem. C 113 (2009) 5700-5704.

[7]  Graham, U.M.; Khatri, R.A.; Dozier, A.; Jacobs, G.; Davis, B.H. Catal. Lett. 132 (2009) 335-341.

[8]  Arai, S.; Muto, S.; Murai, J.; Sasaki, T.; Ukyo, Y.; Kuroda, K.; Saka, H. Mater. Trans. 45 (2004) 2951-2955.

[9]  Arai, S.; Muto, S.; Sasaki, T.; Tatsumi, K.; Ukyo, Y.; Kuroda, K.; Saka, H. Solid State Commun. 135 (2005) 664-667.

[10] Wang, R.; Crozier, P.A.; Sharma, R. J. Mater. Chem.20 (2010) 7497-7505.

[11] Seyedmohammad S.Shams, Rani F. El-Hajjar, Compos. Part A-Appl. S, 49 (2013) 148–156.

[12] Basu, J.; Winterstein, J.P.; Carter, C.B. Appl. Surf. Sci. 256 (2010) 3772-3777.

[13] Avila-Paredes, H.J.; Kim, S. Solid State Ionics 177 (2006) 3075-3080.

[14] Seyedmohammad S.Shams, Rani F. El-Hajjar, Int J Mech Sci, 67 (2013) 70–77.

[15] Li, Z.P.; Mori, T.; Auchterlonie, G.J.; Zou, J.; Drennan, J. Phys. Chem. Chem. Phys. 13 (2011) 9685-9690.

[16] A. Tonomura, Electron Holography, second Ed. Springer, Vol. 70, Springer, 1999.

[17] Tavabi, A. H., Yasenjiang, Z., Tanji, T. J Electron Micros 60 (2011) 307-314.

[18] Tavabi, A. H., Arai, S., Tanji, T. Microsc. Microanal. 18 (2012) 538–544.

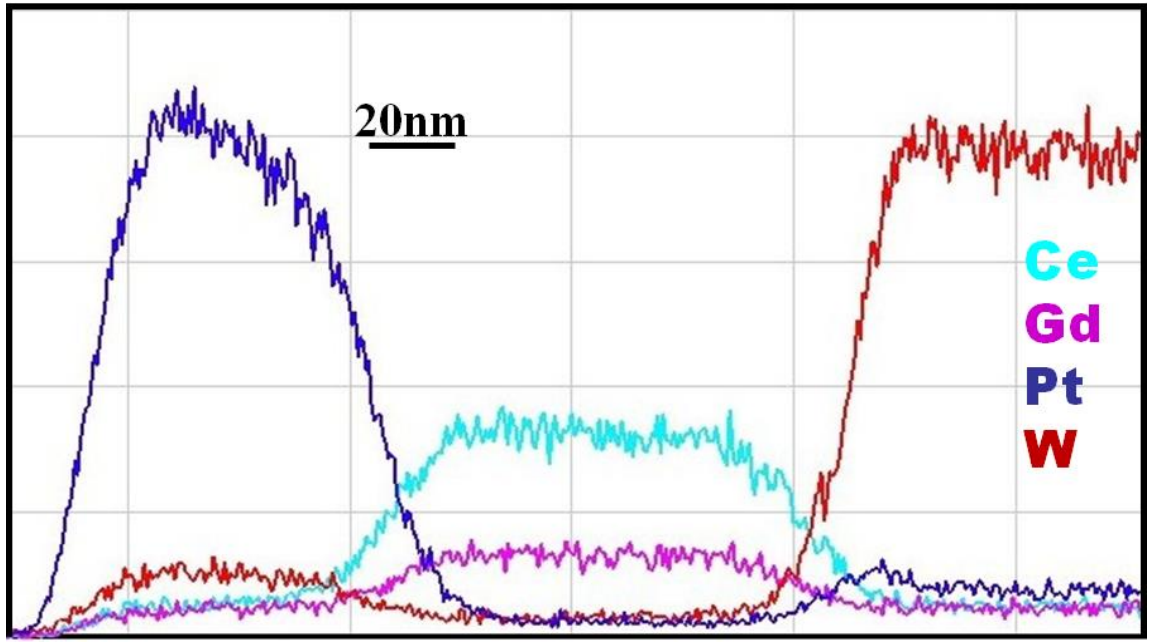

Fig 1. The concentration profile of the EDS line scan obtained perpendicular to the GDC interfaces.

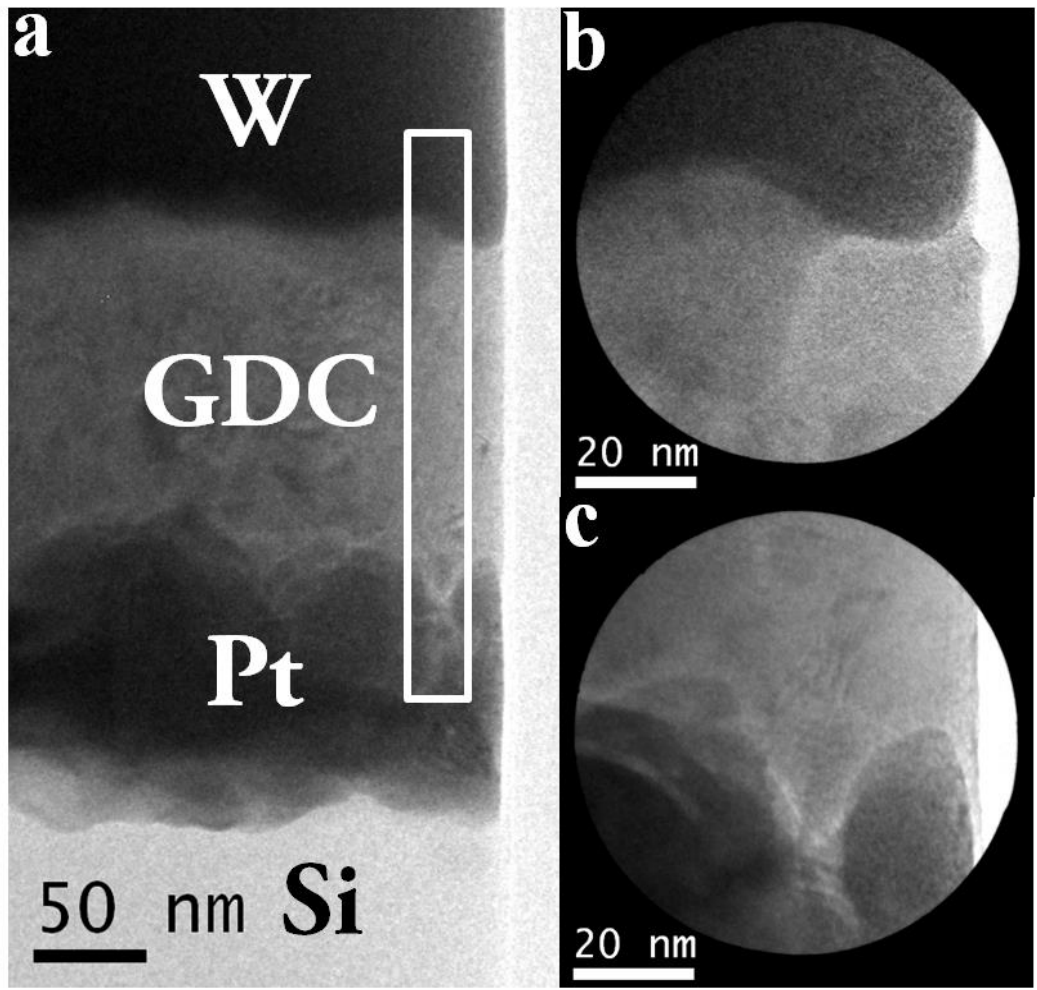

Fig. 2. a) A bright field TEM image of a half cell structure on the Si substrate. The white rectangle shows the area where phase shift profiles were extracted from the phase mages. b) W/GDC interface and c) GDC/Pt interface with the applied apertures for energy loss spectroscopy measurements.

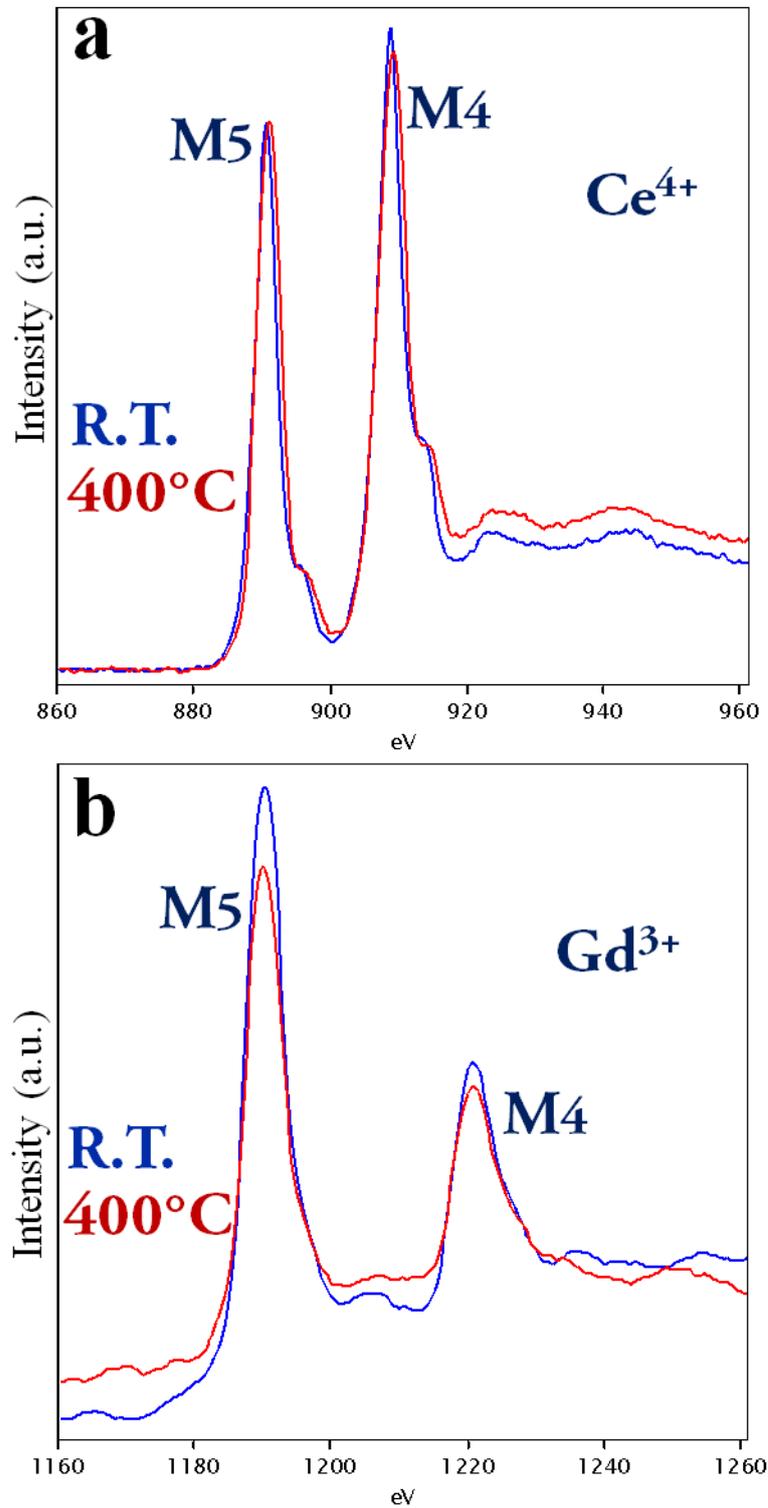

Fig. 3. Energy loss spectra at room temperature and 400°C acquired from the W/GDC interface area for a) cerium and b) gadolinium which show no reaction in this area.

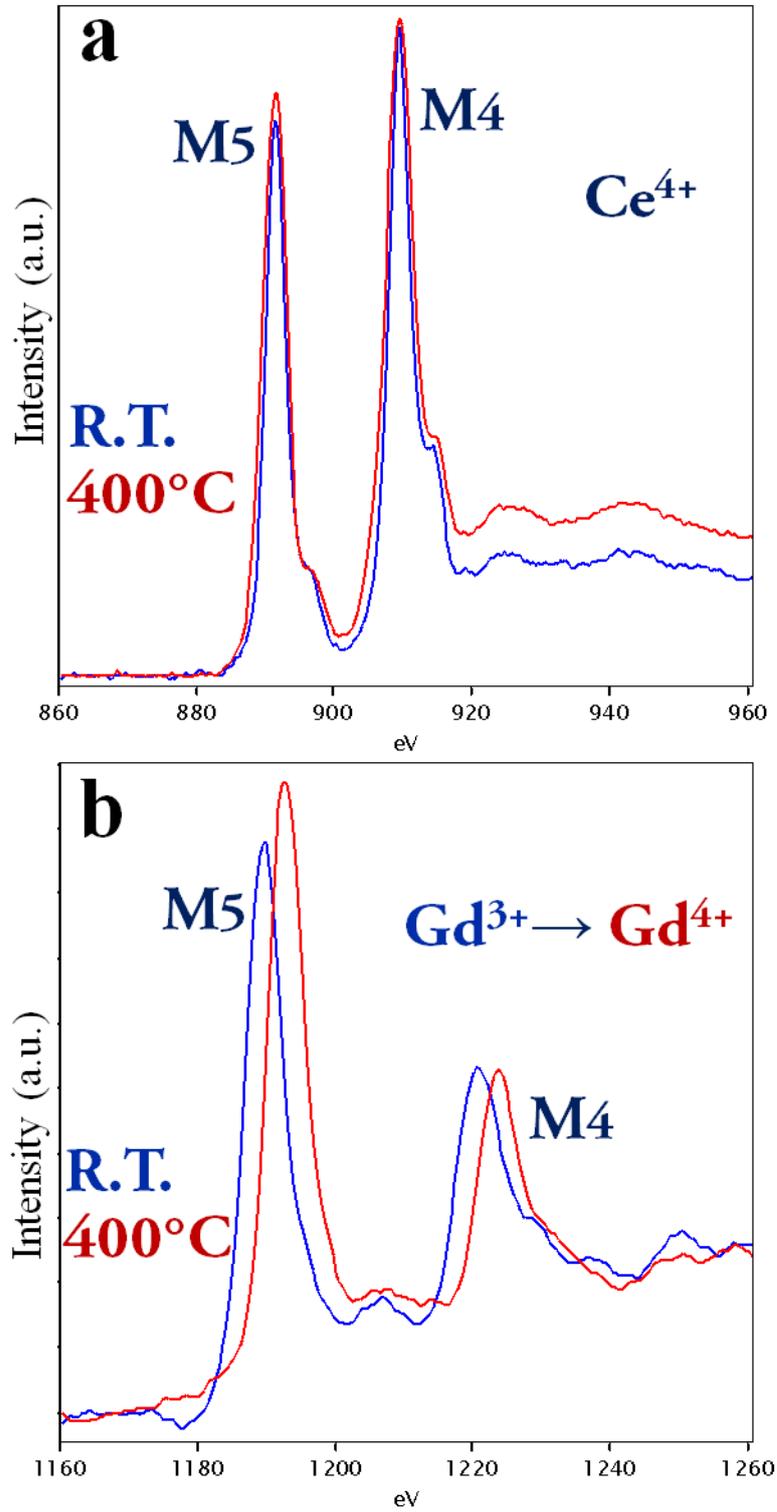

Fig. 4. Energy loss spectra at room temperature and 400°C acquired from the GDC/Pt interface area for a) cerium and b) gadolinium which show a change in gadolinium

oxidation state at high temperature due to oxygen release reaction from this interface.

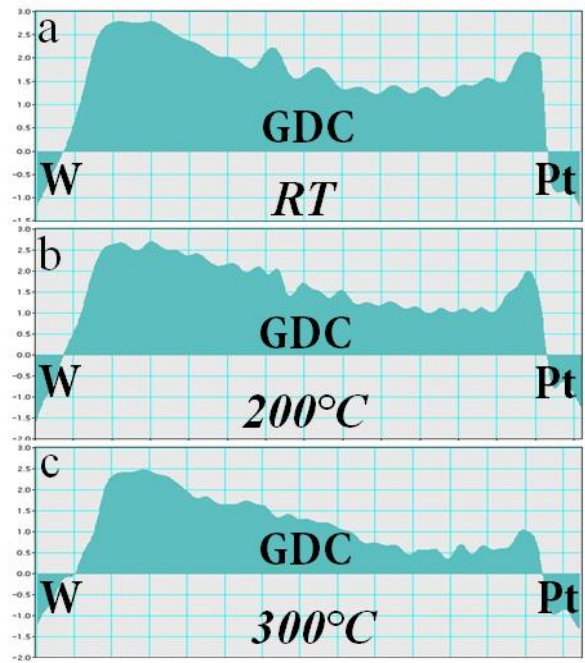

Fig. 5. Phase shift profiles of the half cell at a) room temperature, b) 200°C, and c) 300°C. Phase values of the GDC layer are almost stable at the W interface. However the difference of the Pt and GDC phase amounts decreases as temperature increased.